\documentclass[lettersize,journal]{IEEEtran}

\usepackage[english]{babel}

\usepackage[letterpaper,top=2cm,bottom=2cm,left=2.5cm,right=2.5cm,marginparwidth=1.75cm]{geometry}
\usepackage{xcolor,colortbl}%https://www.overleaf.com/project/626c160c4f5778ede323af95

\definecolor{LightGray}{gray}{0.50}
\definecolor{LightCyan}{rgb}{0.88,1,1}
\usepackage{amsmath}
\usepackage{graphicx}
\usepackage[colorlinks=true, allcolors=blue]{hyperref}
\usepackage{graphicx}
\usepackage{float}
\usepackage{booktabs}
\usepackage{array, caption, tabularx, makecell, booktabs}%
\usepackage{amsmath,physics}
\usepackage{multirow}
\usepackage{xcolor}
\usepackage[linesnumbered,ruled,vlined]{algorithm2e}
\usepackage{soul}
\usepackage[normalem]{ulem}
\usepackage{enumitem}
\usepackage{comment}

\begin{document}

%don't want date printed
\date{}

%make title bold and 14 pt font (Latex default is non-bold, 16 pt)
\title{Navigating the Web of Misinformation: A Framework for Misinformation Domain Detection Using Browser Traffic}

\author{Mayana Pereira, Kevin Greene, Nilima Pisharody, Rahul Dodhia, Jacob N. Shapiro, Juan Lavista

\thanks{M. Pereira, R. Dodhia and J. Lavista are with Microsoft Corporation, AI for Good Research Lab, Redmond, WA (e-mail: mayana.wanderley@microsoft.com).}% <-this % stops a space
\thanks{K. Greene, N. Phisharody and J.N. Shapiro are with Empirical Studies of Conflict Project, Princeton University, Princeton, NJ}
% <-this % stops a space
}


% Remember, if you use this you must call 
%\IEEEpubidadjcol 
%in the second

\maketitle
\begin{abstract}
The proliferation of misinformation and propaganda is a global challenge, with profound effects during major crises such as the COVID-19 pandemic and the Russian invasion of Ukraine. Understanding the spread of misinformation and its social impacts requires identifying the news sources spreading false information. While machine learning (ML) techniques have been proposed to address this issue, ML models have failed to provide an efficient implementation scenario that yields useful results. In prior research, the precision of deployment in real traffic deteriorates significantly, experiencing a decrement up to ten times compared to the results derived from benchmark data sets. Our research addresses this gap by proposing a graph-based approach to capture navigational patterns and generate traffic-based features which are used to train a classification model. These navigational and traffic-based features result in classifiers that present outstanding performance when evaluated against real traffic. Moreover, we also propose graph-based filtering techniques to filter out models to be classified by our framework. These filtering techniques increase the signal-to-noise ratio of the models to be classified, greatly reducing false positives and the computational cost of deploying the model. 
Our proposed framework for the detection of misinformation domains achieves a precision of 0.78 when evaluated in real traffic. This outcome represents an improvement factor of over ten times over those achieved in previous studies. %Our misinformation domain detection pipeline is the first proposal of a practical implementation that can effectively detect misinformation, significantly impacting the awareness of misinformation and its harmful effects on society.

%This paper explores browser traffic data and graph analysis approaches for detecting misinformation websites (sites that publish false or misleading content) using traffic information and host-based features. Our proposed model successfully differentiates misinformation websites from non-misinformation websites. Our deployment framework can discover previously unknown misinformation websites with high precision in real-world scenarios. Our experiments show that the proposed classifier identifies misinformation domains with a precision of 0.98  and a recall of 0.97. When deploying our classifier to browser traffic data, we identified 165 new misinformation domains, which content moderators manually verified. The precision of our classifier in detecting misinformation websites in browser traffic is 0.87. We further extend our model to identify Russian propaganda sites and similarly observe high accuracy and precision. 
\end{abstract}

\section{Introduction}

%%%% elections, public health etc
%The impact of misinformation in modern society is undeniable. In which areas and to which extent misinformation impacts our lives is a question that researchers from a diverse set of disciplines try to answer. The 2016 presidential election in the United States led to a noticeable increase in research on misinformation consumption and its impacts. 

Misinformation has far-reaching implications. Fake stories, such as the `Pizzagate' conspiracy theory, such as \cite{kang2016,grinberg2019fake,lytvynenko2020} have contributed to the erosion of trust in domestic democratic institutions, while other narratives have been deployed to undermine foreign states \cite{mueller2019mueller, zannettou2019disinformation}. In the public health sector, misinformation about potential treatments for COVID-19, such as the promotion of hydroxychloroquine, ivermectin, and disinfectants, resulted in harmful health outcomes and even death \cite{nightingale2020examining}. Further, false claims mislead the population about vaccine safety and efficacy, increasing the challenge for public health officials to reach vaccination targets, likely prolonging the pandemic, and increasing the risk of new virus variants emerging \cite{galhardi2022fake, silva2023covid}.

The expansion of social media and digital platforms allowed false narratives to reach a much larger audience than they would through traditional, analog media. Identifying the channels through which misinformation propagates is a key issue to address. Researchers from diverse fields, such as social sciences, computer science, and various industries – including social media and security – have all devoted efforts to detecting and understanding the online spread of misinformation.

% Why domain-level detection is important?
Previous efforts to combat misinformation have employed strategies such as identifying social media profiles responsible for spreading false information \cite{chen2020proactive}, categorizing news stories as misinformation \cite{perez2017automatic}, and classifying domains as misinformation propagators \cite{hounsel2020identifying, avast, panayotov2022greener}. The latter approach, in particular, has profoundly impacted the detection and containment of misinformation spread. Organizations that concentrate on identifying misinformation domains, such as NewsGuard \cite{newsguard2021} - which maintains a database of domains that fail to adhere to ethical journalistic standards, play a crucial role in informing citizens and media moderators about untrustworthy news sources.

% What the proposals for domains detection do?

%two sentences about describing overall idea of previous approaches 
Domain-level identification proposals have been relatively successful in classifying domains as misinformation/non-misinformation in benchmark data sets. The approaches include domain name and host characteristics analysis \cite{hounsel2020identifying}, social media feeds \cite{chen2020proactive}, a combination of HTML tags and content from a sample of domain URLs \cite{avast}, and Alexa website ranking for creating domain feature sets \cite{panayotov2022greener}.    

However, a challenge highlighted by studies such as \cite{hounsel2020identifying,chen2020proactive} is that models performing well on labeled data sets demonstrate poorer performance when deployed to identify new sites not present in the original training or testing data. One possible reason for this degradation is the scarcity of misinformation sites compared to news sites \cite{hounsel2020identifying}, as supported by previous research revealing that misinformation sites constitute a relatively small proportion of total links shared on social media \cite{guess2019less}. 

%Additionally, during significant events like the COVID-19 pandemic or Russia's invasion of Ukraine, identifying domains spreading misinformation focused on specific narratives becomes crucial for effectively addressing false rumors. Recently, Russia and China created a network of websites to spread propaganda and false narratives. In the case of Russia, of over 60 websites carefully impersonating legitimate European news websites \footnote{https://about.fb.com/news/2022/09/removing-coordinated-inauthentic-behavior-from-china-and-russia/}. To the best of our knowledge, previous efforts have not successfully developed models capable of effectively classifying multiple classes of misinformation.

During significant events, identifying domains spreading misinformation focused on specific narratives is crucial. Recently, Russia and China have created a network of websites to spread propaganda and false narratives, with Russia having over 60 websites carefully impersonating legitimate European news websites, posting articles criticizing Ukraine and  supporting Russia \cite{meta2022}. Previous efforts have not successfully developed models capable of effectively classifying multiple classes of misinformation.

% where they fail?
%% second item has to make it vary clear thet without a deployment framework,
% all proposals seem useless, since it generates a huge burden for content moderators to go over all the produced data.

Thus, we highlight two current challenges in developing effective misinformation domain classifiers:
\begin{enumerate}
%\item Complex feature construction: Constructing model features involves extracting data from various sources, including HTML content \cite{avast, panayotov2022greener} and TLS/SSL certificates \cite{hounsel2020identifying}, as well as processing multiple pages for a single domain \cite{avast}.
\item Decreased performance on new misinformation sites: Many existing approaches experience a drastic decrease in performance when deployed to identify newly emerged misinformation sites \cite{hounsel2020identifying, chen2020proactive}. 

\item Ability to identify domains specialized in specific narratives of misinformation: Previous efforts have yet to successfully develop models capable of classifying multiple classes of misinformation domains that focus on specific narratives. Despite identifying that domains spreading similar narratives link to each other \cite{avast}, there are no proposals for multi-class classification for identifying domains focused on specific narratives.
\end{enumerate}

This paper shows that using information derived from inter-domain traffic patterns significantly improves the current state-of-the-art methodologies. Our proposed framework exhibits an over ten times improvement in precision (0.78 vs 0.05) metrics compared to existing benchmarks for post-deployment performance \cite{hounsel2020identifying}. 

%This paper presents a novel framework for the detection of misinformation domains. Our framework significantly improves the state-of-the-art by increasing precision and recall as well as reducing the deployment's computational costs.

%The central idea of our solution is to use inbound and outbound traffic from domains  to build features that are fed into a machine learning classifier. 

%This paper presents a pipeline that combines training and deployment strategies for effectively classifying disinformation domains and detecting previously unknown instances. 

The underlying insight driving our approach is that navigational patterns of disinformation domains, such as navigations from social media  platforms \cite{fourney2017geographic} and other disinformation domains \cite{greene2023rabbit}, exhibit a distinct nature compared to navigations involving reliable sites. Leveraging this intuition, we adopt a graph-based approach to capture these navigational patterns and create traffic-based features which are fed into a classification model. Despite the abundance of findings pointing to the differences in traffic flows between misinformation and authoritative news sites\cite{greene2023rabbit}, to the best of our knowledge, this insight has not been extensively incorporated into models to identify misinformation domains. 

%Our work focuses on extracting domain features from user navigational patterns associated with misinformation consumption, as noted by \cite{greene2023rabbit}.

%We find that our model is able to identify misinformation domains with high precision (0.99) and recall (0.92). We also train model for multi-class classification (non-misinformation/misinformation/state-sponsored misinformation) and obtain high precision (0.99) and recall (0.97) when classifying state-sponsored misinformation domains. 

By modeling traffic patterns as graphs and graph-related features, we can obtain a performance with actual traffic data that is dramatically superior to previous results in the literature. We verify our model's production predictions with a team of experts in online misinformation. Our deployed model's precision is 0.78 when classifying new misinformation domains, which is over one order of magnitude better than past efforts \cite{hounsel2020identifying}. %We also show that deploying our model to identify state-sponsored misinformation outlets also leads to good results; with a precision of 0.56, we identified 32 new Russian propaganda sites.

Commonly, misinformation classifiers serve as an initial filtering mechanism, flagging potential domains subject to human review. Upon verification, these domains are added to established lists and repositories of misinformation sources. However, when applied to vast data sets, which could encompass hundreds of millions of entries, the prevalent class imbalance issue results in excessive false positives. This significantly hinders the feasibility of the human verification process, essentially overwhelming reviewers with an unmanageable volume of flagged items. Moreover, applying the machine learning classifier to hundreds of millions of domains can also be prohibitively from a computational, economic, and environmental point of view. 

Our paper also presents a solution to this problem. Our proposed framework incorporates a deployment stage where we present a filtering technique to select a subset of domains with a higher signal-to-noise ratio, resulting in a reduced list of flagged domains but with a higher concentration of actual misinformation sources.  By narrowing our focus to these specific sets, we achieve an efficient deployment, requiring performing inference on a significantly reduced number of domains. This approach streamlines the process of identifying new disinformation domains, optimizing  time and computational resources. This is particularly important for non-English websites, given that the vast majority of the fact-checking resources focus on content in English.

\subsection{Related Work}

%%%%  Add more general MI lit 
Understanding the diffusion of misinformation is a topic that drives research in many disciplines \cite{vosoughi2018spread,badawy2019falls, fourney2017geographic, chalkiadakis2021rise}. 
Aside from understanding how widespread the problem is \cite{del2016spreading,allcott2019trends, chalkiadakis2021rise}, researchers have asked questions such as: who are the main spreaders \cite{zubiaga2016analysing, spangher2018analysis}; who are the primary consumers \cite{badawy2019falls}, and can fact-checking help mitigate the harmful impacts of misinformation \cite{dulhanty2019taking,miranda2019automated}.

A focus in recent research has been on employing machine learning techniques to detect domains disseminating misinformation. The approaches to identifying misinformation domains can be broadly classified into two main categories: style-based and propagation-based detection methods, as outlined in a recent survey \cite{zhou2020survey}.

The intuition behind style-based methods is that malicious entities leave distinct markers in the content they create, either because they write fake news in a ``special'' style to increase trust, or simplify their message to increase engagement \cite{zuckerman1981verbal,afroz2012detecting,rubin2015towards,conroy2015automatic,rubin2015deception,rashkin2017truth,potthast2017stylometric,o2018language,horne2017just, chen2020proactive}. %For example, the four-factor model \cite{zuckerman1981verbal} believes that a lie can be characterized by by arousal, attempted control, feelings of guilt, and cognitive effort. 
Research has demonstrated that images associated with fake news tend to exhibit higher levels of clarity and coherence and lower diversity and clustering scores compared to those found in reliable news \cite{7589045}.
Misinformation detection is usually treated as a classification problem, where the underlying text of the sites are candidate features. Some efforts use a hand-crafted feature extractor along with a classifier  \cite{zhou2020fake,oshikawa2018survey}. Others use deep learning based methods such as RNNs and LSTMs to extract features automatically \cite{oshikawa2018survey}. Another relevant line of work in misinformation domain classification captures domain-specific features \cite{hounsel2020identifying}, such as domain name style features, domain host characteristics, and DNS characteristics for misinformation domain classification.

Propagation-based methods assume that the dissemination of fake news is distinguishable from that of authentic news, especially via social bots. Previous work finds that fake news spreads distinctively from real news even at the early stages of propagation \cite{zhao2020fake}. One way to detect fake news is to first identify malicious users \cite{yang2019arming}, or clusters of malicious users that may be acting in coordination \cite{martino2020survey}. Alternatively, the whole propagation tree can be used as input, with features then extracted via pre-defined functions \cite{wu2015false} or learned via neural networks \cite{liu2018early} before applying a classification model.

Often running parallel to efforts to identify misinformation domains, have been studies on how users reach misinformation and of the traffic to and from misinformation sites more generally. For instance, \cite{fourney2017geographic} analyzed Edge browser traffic from 2016 and found out that 68\% of misinformation news traffic comes from social media (Facebook and Twitter), while \cite{spangher2018analysis} found similar results \cite{chalkiadakis2021rise} show that misinformation website traffic comes primarily from direct browser access. Research in the social sciences using browser activity for a representative sample of American users find Facebook and search engines are the most prominent gateways to misinformation \cite{guess2020exposure,guess2020sources,moore2023exposure}. Scholars have noted that misinformation domains appear to exist in their own online ecosystem which is often self-referential \cite{starbird2018ecosystem} - that reputable media outlets are unlikely to share links to known misinformation spreaders \cite{ sehgal2021mutual, greene2023rabbit}. The key point is there appear to be meaningful differences in the traffic flow to and from misinformation domains and authoritative media domains \cite{hanley2021no,sehgal2021mutual}.

These powerful insights are usually not incorporated into models aiming to identify new misinformation domains. The exceptions are \cite{baly2018predicting,avast, panayotov2022greener}.  
\cite{baly2018predicting} uses domain traffic information and while they did not find this feature to improve performance, they suggest that more sophisticated traffic features could lead to better results. \cite{panayotov2022greener} utilizes uses the Alexa web traffic platform to obtain, for each domain, a network of domains that share a similar audience. The domain network in conjunction with graph neural networks are used to classify domains as misinformation/non-misinformation. \cite{avast} extracts embedded hyperlinks within a sample of misinformation and non-misinformation sites. They construct a measure of the proportion of each site's incoming and outgoing links that are known misinformation domains. We build on these efforts by incorporating additional traffic-based features to identify misinformation domains.

\subsection{Summary of Contributions} 

\begin{figure*}
  \centering
 \includegraphics[width=13.3cm]{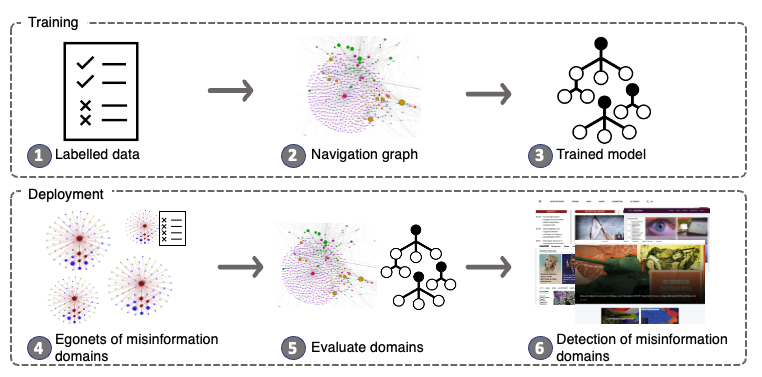}
  \caption{Pipeline for model training for classification of misinformation and propaganda domains. (1) We collect anonymized and aggregated traffic data. (2) We extract traffic-based and host-based features for every domain in our experiments data set. (3) Using our experiments data set, we train several machine learning models for misinformation and propaganda domain classification. (4) During model deployment we select unlabelled domains from ego-nets of misinformation domains to (5) serve as input to our models and (6) identify previously unknown misinformation domains.}  
  \label{fig:training}
\end{figure*}

In this work, we propose a domain classification framework for discovering misinformation domains based on browser traffic patterns (Figure~\ref{fig:training}). Our contributions are summarized as follows:

\begin{itemize}
    \item[1] We develop a method for accurately classifying domains as misinformation/non-misinformation based on a graph representation of browser traffic patterns. 
    \item[2] We show that our method adapts well to a multi-class scenario to find subsets of misinformation domains, such as state-sponsored propaganda.
    \item[3] We propose an effective and practical deployment strategy that filters out domains to be classified based on our graph theoretical modeling of traffic patterns, increasing signal-to-noise ratio of the domains flagged as misinformation and reducing false positives.
    \item[4] We evaluate the efficiency of our proposed framework using a data set collected from real traffic consisting of over 100 Million data entries per month. 
    \item[5] Our implementation achieves a precision for real traffic that is over one order of magnitude greater than previous efforts (0.78 vs 0.05) \cite{hounsel2020identifying}.
\end{itemize}

%We present a framework for efficient deployment for the discovery of misinformation domains. The proposed model and deployment framework are also easily adaptable to subgroups of misinformation, such as propaganda. 

%To the best of our knowledge, our work is the first framework to propose a practical training, testing, and deployment strategy for misinformation domain discovery with high precision rates. Moreover, our work is easily adaptable to multi-class scenarios to find specific subsets of misinformation content.

\section{Overview of Proposed Framework}

In this section, we give an overview of the proposed framework, depicted in Figure \ref{fig:training}. The heart of our framework is a graph representation of the inbound and outbound traffic of internet domains. 

\noindent
\textbf{Navigation Graph} A \emph{navigation graph} $\mathcal{NG}$ consists of a set of nodes $N=\{n_i\}_{i=1}^m$ and a set of directed and weighted edges $E=\{e_{i,j}\}$, where $n_i,n_j \in N$. Nodes are internet domains and a directed edge $e_{i,j}$ connects nodes $n_i$ to $n_j$ if there exists traffic flow from  domain $n_i$ to domain $n_j$. The weight of $e_{i,j}$ is the number of visits to node $n_j$ originating from node $n_i$ for a given period of time (in our case, one month). $\mathcal{NG}$ is a labeled graph where each node can be labeled as a misinformation source or not (binary classification case) or as in multiple misinformation categories and non-misinformation (multi-classification case). 

\noindent
 \textbf{Egonet} Let $\mathcal{NG}$ denote a navigation graph and let $n_i$ be a node in $\mathcal{NG}$. The \emph{one-hop egonet}\cite{akoglu2010oddball} of the node $n_i$ is defined as the induced subgraph of $\mathcal{NG}$ consisting of $n_i$, all nodes directly connected to $n_i$, and all edges between these nodes.

For a general $k$-hop egonet, where $k \geq 2$, we define it as the induced subgraph of $\mathcal{NG}$ that includes $n_i$, all nodes that can be reached from $n_i$ in $k$ or fewer hops, and all edges between these nodes.

We divide our framework in training  and deployment phases. For the sake of simplicity, we present our framework for a binary classification task (misinformation vs non-misinformation). The framework naturally generalizes to a multi-classification setting where we have several classes of misinformation. 

\noindent
\textbf{Training} Our training data set consists of a navigation graph $\mathcal{NG}$ for a period of one month consisting of benchmark domains coming from labeled data sets (see Section \ref{section:dataset}). For each node $n_i$ we extract several features from this graph including traffic from and to authoritative domains, Search Engines, Social Media, News, Email, and other misinformation nodes (see Table \ref{tab:features}). We also collect all the inbound and outbound traffic to and from the egonet defined by $n_i$. 

%Additionally to our graph-based features, we also use a few host-based features associated with the domain corresponding to the node $n_i$, namely its registar, creation date, its registrant country, and DNSSEC extensions used by the domain. 

We use these features to train a machine learning model to classify each node in the data set into a misinformation node or a non-misinformation node. 

\noindent
\textbf{Deployment} For deployment, we assume we have a collection of domains to be classified and their corresponding inbound and outbound traffic.

A direct application of the model would consist of creating the navigation graph $\mathcal{NG}$ corresponding to all the domains available and performing the classification of all of its nodes. 

While such direct application of our framework is interesting in itself, we also propose to build features \emph{only for the domains within the 1-hop egonet of known misinformation domains}. Such a list of misinformation domains can come from the labeled training data set, or from any other curated source. We call this technique \emph{traffic filtering}. It helps with increasing the signal-to-noise ratio in the data, significantly reducing false positives. The proposed filtering mechanism is adaptable and can be tailored to different deployment scenarios, including the identification of specific types of misinformation narratives (e.g., state-sponsored propaganda)

\section{Data sets and features}
\label{section:dataset}

In this section, we give a detailed description of our data and domain features used in our misinformation detection models. We work with novel  traffic features combined with previously proposed host-based features\cite{hounsel2020identifying}.% We explore the scenario where only traffic-based features are utilized for model training and compare it with models trained on the combination of  traffic and host-based features.
\subsection{Data sets}

The datasets described below are used to train and test the machine learning models described in section \ref{sec:model}.

\subsubsection{NewsGuard data set}

NewsGuard \cite{newsguard2021} is a journalism and technology company that rates the credibility of news websites. NewsGuard scores domains based on nine apolitical criteria that assess basic practices of credibility and transparency.\footnote{The NewsGuard data were downloaded on December 6, 2022.} Domains that receive a score below 60 out of 100 fail to meet basic standards of credibility and transparency according to NewsGuard and are treated as misinformation domains for our evaluations, while domains with a score greater than 60 are classified as non-misinformation domains.\footnote{The domains labeled by NewsGuard as non-misinformation are also referred to as \emph{authoritative news domains} throughout this paper.} Both domain types are used in our experiments. 

\subsubsection{GDI data set}

The Global misinformation Index (GDI) \cite{srinivasan2019rating} is a non-profit organization that produces independent ratings of news sites’ misinformation risk. The risk ratings are then used by advertising technology companies to ensure sites do not support high-risk websites. The dataset provided by GDI used in our experiments contains 1436 domains that were manually verified as misinformation spreaders.\footnote{The GDI data were downloaded on December 6, 2022.}

\subsubsection{Propaganda domains list} 

Strategic analysts specializing in the detection and countering of malign influence and extremism from a company formerly known as Miburo\footnote{Miburo was recently acquired by Microsoft. https://miburo.substack.com}, compiled a list containing 185 domains that were verified as Russian propaganda spreaders.\footnote{The data were downloaded on December 6, 2022.} 

\subsubsection{Edge logs data set}
\label{sec:edge}

We collected three months of instrumentation data from Edge, a desktop web browser. All data was collected from opt-in users, and all personal identifying information was removed. Our analysis begins on January 1, 2022, and ends on March 31, 2022. Along with the lists of domains provided by GDI and NewsGuard, we also queried the top 4,000 domains, ranked by unique visits in the Edge Browser. %We analyze the 5k domains and observe that 26 domains are labeled as misinformation by either NewsGuard or GDI. 
Of these 4,000 the 3,976 domains not contained in existing lists of misinformation domains serve as part of the non-misinformation domains dataset. 

Subsequently, we collected traffic information. The traffic dataset consists of timestamped inbound and outbound traffic information for each domain in our analyses. Inbound traffic consists of the domain visited before the domain of interest. Outbound traffic consists of the domain visited after the domain of interest. For our study, we collected visits to all domains contained in the datasets listed in Table \ref{tab:widgets}, which includes misinformation domains and non-misinformation domains. Data download date: December 6, 2022.

\begin{table*}
\centering
\begin{tabular}{l c c c}
 \hline
Source & misinformation & Non-misinformation & Propaganda\\\hline
NewsGuard & 3343  & 5116  & 44\\
GDI & 2755 & - & 135 \\
Propaganda Domains List & 186 & - & 186\\
Top Domains & 25 & 3974 & 1 \\
\hline
Unique domains & 5558 & 8874 & 186

\end{tabular}
\caption{\label{tab:widgets} Our Experimental Dataset. The experimental dataset consists of misinformation domains from NewsGuard and GDI, non-misinformation domains from NewsGuard, and popular domains from Edge. Note that the count of unique domains does not match the sum of the domains as several domains appear in multiple data sets. Propaganda domains are a subset of misinformation domains. }
\end{table*}

\subsection{Model Features}\label{sec:feat}

We now give a more detailed explanation of the features used in our proposed framework. Our framework uses solely traffic-based features extracted from the corresponding navigation graph. However, we include some analysis of host-based features used in past work \cite{guess2020exposure,guess2020sources,hounsel2020identifying}. Our results show that host-based features do not increase the performance of classifiers trained solely on traffic-based features. 

\subsubsection{Traffic features} We use the Edge Logs dataset to construct navigation graph $\mathcal{NG}$. In $\mathcal{NG}$, each node is a domain, and a directed edge from node $n_i$ to node $n_j$ exists if there were at least 3,000 visits to domain $n_i$ after being referred from domain $n_j$ (or vice-versa) in one month. We obtain this information by checking the HTTP referrer. The HTTP referrer is an optional HTTP header field that identifies the address of the web page from which the current page has been requested.

To create traffic features for each domain we use egonets. We define two egonets for every domain in our data set. The outbound egonet of a domain is the 1-hop subnetwork that represents all traffic for which the domain is a source of traffic (a referrer). Analogously, we define a domain's inbound egonet as the 1-hop subnetwork that represents all traffic for which the domain is a traffic target.

\begin{figure}
  \centering
 \includegraphics[width= 6cm]{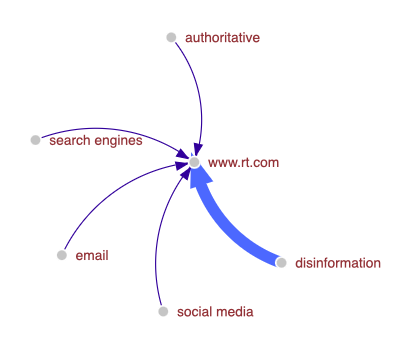}
  \caption{Simplified representation of an inbound egonet of a misinformation domain. The thicker the edge between two nodes, the more traffic flows between the sites.} \label{img:inbound}
\end{figure}

For every domain in $\mathcal{NG}$, we identify its outbound and inbound egonets, and extract domain traffic features, as shown in Figure \ref{img:inbound}. 

\begin{figure*}
  \centering
 
 \includegraphics[width=15cm]{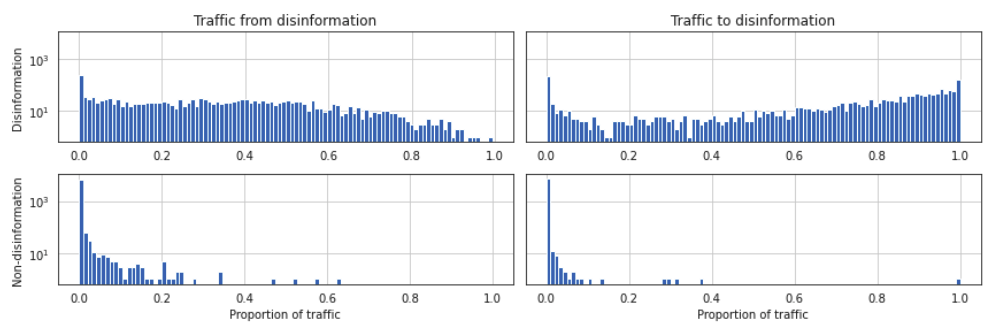}
  \caption{Differences in the proportion of traffic to and from misinformation domains for misinformation and non-misinformation domains. The histogram displays the number of domains based on the proportion of traffic going to or coming from other misinformation domains.} \label{img:traffic}
  
\end{figure*}

% \mayana{Nilima and Kevin, could you help me making sure we do not forget any citations here? }
The choice of model features comes from insights from existing literature. Research conducted by \cite{hanley2021no, sehgal2021mutual} has revealed a significant concentration of traffic within misinformation domains, with these domains primarily interacting among themselves rather than linking to reputable sites \cite{greene2023rabbit}. 
We also have verified a similar pattern in our data. In figure \ref{img:traffic} we observe clear differences in the traffic patterns of misinformation and non-misinformation domains. The majority of non-misinformation domains have almost no traffic coming from or going to misinformation domains. However, much of the outbound traffic from misinformation domains is sent to other misinformation domains.

%Previous work \cite{hanley2021no,sehgal2021mutual} has found that misinformation domains heavily dominate the traffic to and from other misinformation domains, while rarely being linked to by reputable sites \cite{greene2023rabbit}. For each domain, we calculate the total traffic coming from previously identified misinformation and reputable sites. 

We include features that measure the total traffic from Google, Bing, and DuckDuckGo, since there is significant evidence that search engines are often the largest referrers to misinformation domains \cite{spangher2018analysis,guess2020sources,guess2020exposure, chalkiadakis2021rise,moore2023exposure}. \cite{thompson2022fed} notes that DuckDuckGo, in particular, is increasingly popular among conspiracy theorists. 

%We included features that measure the total traffic from Google, Bing, and DuckDuckGo. 

We also include a feature that measures the total traffic from social media. The social media platforms utilized in our work are \emph{Facebook, Twitter, TikTok, LinkedIn, Telegram}, and \emph{WhatsApp web}.  This choice is supported by previous research that demonstrates social media's tendency to be a source of referral traffic to misinformation sites \cite{fourney2017geographic,guess2020sources,guess2020exposure,chalkiadakis2021rise}.

In addition to the aforementioned features, we also measure traffic from and to news aggregators (\emph{bloomberg, MSN, news.google} and \emph{news.yahoo}); traffic from and to email providers (\emph{gmail, mail.yahoo} and \emph{outlook}); total inbound and outbound traffic.  

Traffic features are outlined in Table \ref{tab:features}. Edge weights are represented by the sum of page views accumulated over one month. For greater clarity, consider the following example regarding feature computation. The feature \emph{Traffic to misinformation} for the domain \emph{www.rt.com} indicates the count of all page visits directed towards misinformation domains with the HTTP referrer identified as \emph{www.rt.com}. For normalization, these counts are adjusted relative to \emph{Inbound Traffic} for inbound features and \emph{Outbound Traffic} for outbound features.

\subsubsection{Host-Based features}

Our proposed framework only uses traffic-based features. However, we present experiments with host-based features previously used in the literature\cite{hounsel2020identifying} to show that they provide no additional predictive power to our set of traffic features. These features describe the properties of the website host. For each site, we collected the registrar name, registrant country, creation date, and the type of DNS security extensions used by the domain. These features allow us to understand the management and maintenance of domains.

%Our experiments also include host-based, such as the ones proposed in \cite{hounsel2020identifying}. The purpose of adding host-based features is to understand if they provide additional predictive power. These features describe the properties of the website host.  For each site, we collected the registrar name, registrant country, creation date, and the type of DNS security extensions used by the domain. These features allow us to understand the management and maintenance of domains.

\begin{table*}
\centering
\begin{tabular}{l l l l}
Feature & Category & Description & Data type\\\hline
   \multicolumn{4}{l}{Traffic features}\\ \hline
Inbound traffic & Traffic & Total Monthly page views referred & Numeric\\
& &  by other domains (log) &\\
Outbound traffic & Traffic & Total Monthly page views referred & Numeric\\
& &  to other domains (log) &\\
Traffic to misinformation & Traffic  & Monthly page views that preceded a visit to & Numeric\\
 &   & a misinformation domain (normalized)  & \\

Traffic to authoritative & Traffic & Monthly page views that preceded a visit to & Numeric \\
 &   & an authoritative domain (normalized)& \\

Traffic to Russian propaganda* & Traffic & Monthly page views that preceded a visit to & Numeric \\
 &   & is a Russian propaganda domain (normalized)  & \\

Traffic to Google & Traffic & Monthly page views that preceded a visit to & Numeric\\
 &   & Google (normalized)  & \\

Traffic to Bing & Traffic & Monthly page views that preceded a visit to & Numeric \\
 &   & Bing (normalized)  & \\

Traffic to DuckDuckGo & Traffic & Monthly page views that preceded a visit to & Numeric \\
 &   & DuckDuckGo (normalized)  & \\

Traffic to social media & Traffic & Monthly page views that preceded a visit to & Numeric \\
 &   & social media (normalized)  & \\

Traffic to news aggregators & Traffic & Monthly page views that preceded a visit to & Numeric \\
 &   & news aggregators (normalized)  & \\

Traffic to mail & Traffic & Monthly page views that preceded a visit to & Numeric \\
 &   & an email provider (normalized)  & \\

Traffic from misinformation & Traffic  & Monthly page views that succeeded a visit  & Numeric\\
 &   & to misinformation domains (normalized)  & \\

Traffic from authoritative & Traffic & Monthly page views that succeeded a visit & Numeric \\
 &   & to authoritative domains (normalized)  & \\

Traffic from Russian propaganda* & Traffic & Monthly page views that succeeded a visit & Numeric \\
 &   & to Russian propaganda domains (normalized)  & \\

Traffic from Google & Traffic & Monthly page views that succeeded a visit & Numeric\\
 &   & to Google (normalized)  & \\

Traffic from Bing & Traffic & Monthly page views that succeeded a visit & Numeric \\
 &   & to Bing (normalized)  & \\

Traffic from DuckDuckGo & Traffic & Monthly page views that succeeded a visit & Numeric \\
 &   & to DuckDuckGo (normalized)  & \\

Traffic from social media & Traffic & Monthly page views that succeeded a visit & Numeric \\
 &   & to social media (normalized)  & \\

Traffic from news aggregators & Traffic & Monthly page views that succeeded a visit & Numeric \\
 &   & to news aggregators (normalized)  & \\

Traffic from mail & Traffic & Monthly page views that succeeded a visit & Numeric \\
 &   & an email provider (normalized)  & \\

Inbound egonets & Traffic & Number of unique misinformation domain & Numeric \\
 &   & inbound egonets the domain is part of  & \\
 Outbound egonets & Traffic & Number of unique misinformation domain & Numeric \\
 &   & outbound egonets the domain is part of  & \\\hline
   \multicolumn{4}{l}{Host features}\\ \hline

Registrar & Host & Organization that manages the reservation & Categorical\\
 &   & of the domain name& \\
Creation date & Host & Date of creation of the domain (year only) & Categorical\\
Registrant country & Host & The country of the registrant & Categorical\\
DNSSEC & Host & DNS security extensions used by the domain & Categorical\\ \hline
\end{tabular}
\caption{Traffic and host features used to train our misinformation domain classifier. Our traffic features are based on traffic patterns from one month of browser traffic. Our list of features includes inbound and outbound traffic to search engines, social media platforms, news platforms, authoritative news websites, and misinformation websites. Our host-based features include registrar information, country, year of creation, and the security extensions used by the domain (if any). Features \emph{Traffic from Russian propaganda} and \emph{Traffic to Russia Propaganda} are only utilized in our multi-class model.}\label{tab:features} 
\end{table*}

\section{Model Training and Evaluation}
In this section, we describe the procedures used for model training and discuss model evaluation.

\subsection{Binary Misinformation Domain Classification}\label{sec:model}

We initially trained machine learning models to classify domains as either misinformation/non-misinformation based on the features described in section \ref{sec:feat}. We utilized three months of Edge logs data (see section \ref{sec:edge}) and divide the data into three distinct monthly data sets: October, November, and December. For each month of data, we created training and testing data sets consisting of the labeled misinformation and non-misinformation domains from table \ref{tab:widgets}, and our traffic-based and host-based features. %The goal is to train our model in on the first month (October), and test the model in all three months to understand how the model behaves over time.

\subsubsection{Model Training and Evaluation}\label{train:binary}

We partitioned the domains into 5-folds to perform a 5-fold cross-validation. In each cross-validation round, we train our model on the 80 split with traffic features from the month of October, and we test the model on the 20 split with traffic features from the months of October, November, and December. We compare the performance of 5 distinct machine-learning techniques. We chose a set of standard ML algorithms with distinct functional forms: K nearest neighbors (KNN), support vector machines (SVM), boosted decision trees (XGB), logistic regression, and random forests. 

We chose these machine learning models because of their interpretability, simplicity, and previous use in the literature. We have avoided using any non-interpretable model to ensure that the models' decisions can be understood, validated, and trusted. 

%\mayana{Include a justification for the choice of the ML techniques (interpretability, simplicity, etc.}

For each machine learning technique, we trained two distinct models: one that takes as input traffic-based features only, and another model that takes as input traffic and host-based features. The purpose is to understand the gain in model performance when utilizing such features.% since host-based features are sometimes difficult to get due to privacy regulations and missing data.

All experiments used the scikit-learn python library. The KNN algorithm uses $k=5$. The SVM uses an RBF kernel. The XGB model uses $max\_depth = 6$. Our logistic regression models used a regularization parameter $C = 1$. Our random forests models had $max\_depth = 20$. We selected as evaluation metrics: accuracy, precision, and recall. The performance details of all our experiments are in Table \ref{model:binary}. All metrics represent the average across 5-folds.

\begin{table*}
\centering
\begin{tabular}{ l c c c c c c c c c }
&\multicolumn{3}{c}{October}&\multicolumn{3}{c}{November}&\multicolumn{3}{c}{December}\\
\hline
\multicolumn{10}{l}{Traffic features}\\ \hline
 {Model} &  Accuracy & Precision & Recall & Accuracy & Precision & Recall   & Accuracy & Precision & Recall  \\\hline
KNN	&	0.96 & 0.93 & 0.88 & 0.96 & 0.94 & 0.88 & 0.96 & 0.93 & 0.87\\
	SVM & 	0.97 & 0.99 & 0.88 & 0.97 & 0.99 & 0.89 & 0.97 & 0.99 & 0.88\\
XGB	&	0.98 & 0.99 & 0.92 & 0.98 & 0.99 & 0.92 &0.98 & 0.99 & 0.92\\
Logistic regression &	0.97 & 0.99 & 0.88 & 0.97 & 0.99 & 0.89 & 0.97 & 0.99 & 0.89\\
	Random forest &  0.98 & 0.99 & 0.91 & 0.98 & 0.99 & 0.92 & 0.98 & 0.99 & 0.92\\

\hline
\multicolumn{10}{l}{Traffic features and host features }\\ \hline
 {Model} &  Accuracy & Precision & Recall & Accuracy & Precision & Recall   & Accuracy & Precision & Recall\\\hline

KNN	&	0.95 & 0.92 & 0.85 & 0.95 & 0.91 & 0.87 & 0.96 & 0.94 & 0.89\\
	SVM & 	0.97 & 0.99 & 0.89 & 0.97 & 0.98 & 0.9 & 0.97 & 0.99 & 0.89\\
XGB	&	0.98 & 0.99 & 0.92 & 0.98 & 0.99 & 0.92 &0.98 & 0.99 & 0.92\\
Logistic regression &	0.97 & 0.99 & 0.88 & 0.97 & 0.99 & 0.89 & 0.97 & 0.99 & 0.88\\
	Random forest &  0.98 & 0.99 & 0.91 & 0.98 & 0.99 & 0.91 & 0.98 & 0.99 & 0.91\\

\hline

\end{tabular}
\caption{ Performance metrics for our classifiers. Model performance evaluation using accuracy, precision, and recall. To evaluate how well the model generalizes to multiple months, the model was trained on data from October and was tested on data from October, November, and December.}\label{model:binary}
\end{table*}

We show our classifiers' accuracy, precision, and recall for October, November, and December in Table  \ref{model:binary}. Our models have similar performance when using traffic and host features and only traffic features. Thus, we can safely remove host-based features without affecting the performance of our classifiers. 

%, leaving us to conclude that traffic features are very strong predictors of whether a domain is a misinformation domain or not, and host-based features do not add any predictive value. 
We further observe that random forests and XGB perform better than logistic regression, KNN, and SVMs. The random forest models have an average precision of 0.98 and an average recall of 0.91 for all three months, and XGB has an average precision of 0.99 and an average recall of 0.92 for all three months. A highlight of our model evaluation is the consistency across all three months. This means that models trained in one specific month can classify misinformation domains in subsequent months without performance decays and potentially require less model retraining. 

Compared to previous efforts to classify misinformation domains evaluated \emph{using benchmark data sets}, the performance of our model surpasses \cite{hounsel2020identifying, panayotov2022greener, chen2020proactive}. Our model presents results comparable to \cite{avast} (ours being slightly better). In Section \ref{sec:real} we show that the  performance of our proposal remains good even when evaluated against real traffic.

%An additional advantage of our traffic-based approach is that traffic-based features are harder to be manipulated by ill-intended actors. However, the true benefit of our traffic-based features will show up when evaluated against reall traffic - a point where previous results showed a poor performance \cite{hounsel2020identifying}. 

%Robustness against model evasion is an important attribute of misinformation domain classification systems. 
%While previous work has noted the importance of robustness against model evasion, features such as meta tags and page content \cite{avast} and domain names \cite{hounsel2020identifying}, which are relatively easy to manipulate, are often used as predictive features.
%%%% Revised up to this point 06/11 8:00am

\subsection{Multi-class Misinformation Domains Classification -  Detecting Russian Propaganda}\label{sec:multiclass}

In this section, we extend our results to a multi-classification setting. We will train machine learning models to classify a domain into three categories: non-misinformation, general misinformation, and Russian state-sponsored propaganda. Russian state-sponsored misinformation was frequent in the run-up to the 2016 US election and the Russian invasion of Ukraine \cite{pierri2023propaganda, badawy2018analyzing}.  To this end, we obtained a list of domains identified by experts to have contributed to previous Russian influence operations. We utilized this list to create a third class of domains and now treat the prediction task as a multi-class classification problem. 

As in section \ref{train:binary}, we collected three months of browser logs data and divided the data into three monthly data sets. For each month of data, we built a dataset consisting of misinformation, Russian propaganda, and non-misinformation domains (Table \ref{tab:widgets}), and for each domain, we extracted traffic-based features and host-based features. We utilized the traffic-based features listed in \ref{tab:features}, including the features \emph{Traffic from Russian Propaganda} and \emph{Traffic to Russian Propaganda}.

\subsubsection{Multi-class Model Training and Evaluation}

As in the binary case, our multi-class experiments use 5-fold cross-validation. In each cross-validation round, for each domain in the testing fold, we evaluate the model three times: first using October data, next using November, and a third time using December data.

We compare the performance of 5 distinct machine learning techniques: K nearest neighbors (KNN), support vector machines (SVM), boosted decision trees (XGB), logistic regression, and random forests. In our multi-class experiments, we used the same hyper-parameters as those defined in section \ref{train:binary}.

%\begin{table*}
%\centering
%\begin{tabular}{l|c|c|c|c|c|c}
%\hline
%&\multicolumn{3}{c|}{Logistic Regression %}&\multicolumn{3}{c}{Random Forests }\\
%\hline
% {Evaluation} &  Accuracy & Precision & Recall & Accuracy & %Precision & Recall \\\hline

%January	&0.99&	0.99&	0.99&	0.99& 0.99 & 0.99\\
%February&	0.99&	0.99 &	0.99 &	0.99 & 0.99 & 0.99\\
%March&	0.99 &	0.99 & 0.99 &	0.99 & 0.99 & 0.99\\

%\hline
%\end{tabular}
%\caption{\label{tab:widgets}Our Experimental Dataset. The %experimental dataset consists of misinformation domains from %NewsGuard and GDI, and non-misinformation domains from %NewsGuard and popular domains from Edge.}
%\end{table*}

We evaluated our multi-class models using accuracy, precision, and recall metrics. Our results are shown in Table \ref{model:propaganda}. The average accuracy, precision, and recall across months for our best-performing model are 0.98, 0.99, and 0.96 respectively. We report precision and recall for the `propaganda' class, which is our class of interest.

\begin{table*}
\centering
\begin{tabular}{ l c c c c c c c c c }
&\multicolumn{3}{c}{October}&\multicolumn{3}{c}{November}&\multicolumn{3}{c}{December}\\
\hline
\multicolumn{10}{l}{Traffic features}\\ \hline

 {Model} &Accuracy & Precision & Recall & Accuracy & Precision & Recall   & Accuracy & Precision & Recall \\\hline
KNN	&	0.97 & 0.94 & 0.88 & 0.96 & 0.92 & 0.84 & 0.97 & 0.94 & 0.85\\
	SVM & 	0.97 & 1.0 & 0.88 & 0.97 & 1.0 & 0.87 & 0.97 & 1.0 & 0.87\\
XGB	&	0.98 & 1.0 & 0.96 & 0.98 & 0.99 & 0.97 &0.98 & 1.0 & 0.96\\
Logistic regression &	0.97 & 0.99 & 0.89 & 0.97 & 0.98 & 0.88 & 0.97 & 0.98 & 0.88\\
	Random forest &  0.98 & 1.0 & 0.96 & 0.98 & 0.99 & 0.97 & 0.98 & 1.0 & 0.96\\

\hline
\multicolumn{10}{l}{Traffic features and host features }\\ \hline
 {Model} &  Accuracy & Precision & Recall & Accuracy & Precision & Recall   & Accuracy & Precision & Recall \\\hline

KNN	&	0.95 & 0.92 & 0.85 & 0.95 & 0.91 & 0.87 & 0.95 & 0.92 & 0.85\\
	SVM & 	0.97 & 0.99 & 0.89 & 0.97 & 0.99 & 0.88 & 0.97 & 0.99 & 0.88\\
XGB	&	0.98 & 1.0 & 0.96 & 0.98 & 0.99 & 0.97 &0.98 & 1.0 & 0.96\\
Logistic regression &	0.97 & 0.98 & 0.89 & 0.97 & 0.97 & 0.89 & 0.97 & 0.98 & 0.88\\
	Random forest &  0.98 & 1.0 & 0.97 & 0.98 & 0.99 & 0.97 & 0.98 & 1.0 & 0.96\\

\hline

\end{tabular}
\caption{ Performance metrics for our multi-class classifiers. Model performance evaluation using accuracy, precision, and recall. We report precision and recall for the `propaganda' class, which is our class of interest. The model was trained using October data. To evaluate how well the model generalizes to multiple months, the model was tested using October, November, and December data.}\label{model:propaganda}
\end{table*}

Once again, XGB and random forest are the best performing models. As with the binary version of the model, adding host-based features does not increase model performance. %These results confirm that the traffic-based features which were powerful for differentiating misinformation domains from non-misinformation domains can also be used to identify specific subgroups of misinformation.

From table \ref{model:propaganda} we observed that in some rounds of model evaluation, the precision is equal to 1, suggesting there are unique traffic patterns that help the model identify Russian propaganda domains. 

%In section \ref{find:russian}, we evaluate web traffic data distinct from the data used to train or test the model for a more accurate understanding of the model's performance and ability to identify new misinformation domains.

\subsection{Feature Analysis}

In this section, we evaluate the features that provide the most predictive leverage for our models, with the goal of better understanding why our models are high performing. We use the feature importance measure for our random forest binary and multi-class models. In table \ref{tab:featureImportance} we present the top 5 features for each of our four types of experiments: binary model with traffic-based features; binary model with traffic and host-based features; multi-class model with binary features; multi-class model with traffic and host-based features.

\begin{table*}
\centering
\begin{tabular}{l l l l l }

 & \multicolumn{2}{l}{Binary model }& \multicolumn{2}{l}{Multiclass model }  \\\hline
Ranking & Feature & Gini importance & Feature & Gini importance\\\hline
1 & Traffic to misinformation	& 0.316 &
Traffic to misinformation	& 0.314 \\

2 & Traffic from misinformation &	0.311 & Traffic from misinformation &	0.266 
\\
3 & Inbound egonets	& 0.071 &
Outbound traffic	& 0.067  \\

4 & Outbound traffic	& 0.043 &
Traffic from authoritative	& 0.065 
\\
5 & Traffic from authoritative &	0.038 &
Traffic to authoritative &	0.057 \\\hline

 \end{tabular}
\caption{Feature importance for our random forests model. We measure feature contribution using Gini importance. As noted in previous works, traffic patterns from and to misinformation are distinct for misinformation and non-misinformation domains, and when utilized as model features for misinformation classification they are powerful predictors.}\label{tab:featureImportance} 
\end{table*}

We used Gini importance to rank the model features. The list of top features confirms what previous works in social sciences \cite{greene2023rabbit} had already hinted: traffic from and to other misinformation domains are excellent predictors of whether a domain is misinformation or not. We utilize this same intuition to propose an efficient model deployment framework. We discuss details of our deployment framework and deployment experiments next.

\section{Model Deployment in Real Traffic}
\label{sec:real}

In this section, we analyze the performance of our framework compared to the current state-of-the-art methods \cite{hounsel2020identifying},  when applied to domains coming from real traffic rather than from benchmark data sets. We perform such analysis by deploying the model in real traffic and manually checking the model's classification results for a random subset of the domains classified.  

%This is achieved by operating our framework on domains found in actual traffic data and manually labeling a randomly chosen subset of these domains. The labels generated through this process are then compared to the outputs produced by our classifier.

We further optimize our methodology by introducing filtering techniques, which refine the pool of domains presented to the classifier. This strategic approach reduces the computational costs of deploying our solution and enhances the signal-to-noise ratio among flagged domains, significantly reducing false positives.

Machine learning models can be used to perform a preliminary selection of domains, flagging those more likely to be misinformation sources. This step is followed by a human review of the flagged models, resulting in a final, curated, list of misinformation sources that is then added to well-known repositories. 

Applying our classifiers directly (described in Sections \ref{sec:model} and \ref{sec:multiclass}) to large data sets would result in a number of flagged domains that would be unreasonably large for posterior human verification. In the case of our previously described models and the data set we use for real traffic validations ( 350 million domains), we end up with  about 800k domains flagged by our classifier, making the human verification process impossible. Furthermore, the computational cost of running the classifier against data sets consisting of hundreds of millions of domains is significant. 

We propose a novel framework for significantly reducing the set of domains presented to human reviewers. We leverage the insights obtained from previous works in social sciences \cite{greene2023rabbit} to increase the signal-to-noise ratio in deployment. We illustrate our pipeline in Figure \ref{fig:training}.

We apply our classifiers only for the domains within the 1-hop egonet of known misinformation domains. Such a list of known misinformation domains can be provided by existing curated lists provided by organizations such as NewsGuard and GDI datasets. 

By applying such a filtering technique, we increase the signal-to-noise ratio of our final list of flagged domains, and reduce its size, while still being able to detect previously unknown misinformation domains. 

%This is the first work that addresses the issue of practical model deployment for misinformation domain identification.

%\ref{sec:model} and \ref{sec:multiclass} aim to create a list of domains that are highly likely to be misinformation or Russian propaganda spreaders. Content rating organizations such as NewsGuard, and social media platforms constantly look for new misinformation domains to track potentially harmful content. Keeping track of the ever-evolving misinformation ecosystem is time-consuming and expensive. New content creators and spreaders frequently appear, and each geolocation/region has its particular ecosystem. Automatized systems for misinformation domain discovery can reduce the time needed to find new misinformation and propaganda domains by several orders of magnitude. The goal is to keep humans in the loop by sending a curated list of potential misinformation domains for human reviewers to evaluate.  

%We propose an effective pipeline for identifying new misinformation and propaganda domains in browser traffic. We leverage the insights obtained from previous works in social sciences \cite{greene2023rabbit} to increase the signal-to-noise ratio in deployment. We illustrate our pipeline in Figure\label{img:train}.

%This is the first work that addresses the issue of practical model deployment for misinformation domain identification.

\subsection{Deployment Framework}

Our deployment framework aims to enhance the signal-to-noise ratio during deployment. With over 350 million domain name registrations \cite{verisign} in our sample, it is crucial to understand how to select domains for use as model input in the deployment pipeline, to improve model performance and ensure human evaluators are not overwhelmed with false positives. 

Our proposed framework is as follows: 

\begin{itemize} 

\item We select domains for the model inference. As shown in Figure \ref{fig:training}, we start by collecting traffic data, specifically from all known misinformation domains. 

\item We generate inbound and outbound egonets for each of these known misinformation domains. Domains (nodes) in egonets of misinformation domains consist of three types: known misinformation, known non-misinformation, and unlabelled domains. 

\item For each unlabeled domain to be classified, we extract traffic features as described in section \ref{sec:feat}. We then feed these features to the trained machine-learning model for classification.

\item Domains identified as misinformation are sent to experts for review, and those confirmed as true positives are added to the labeled domains list. As a result, the number of misinformation egonets increases over time, enhancing the chances of discovering new domains.

\end{itemize}

This framework also generalizes for considering two, or more hop egonets centered in known misinformation domains. The larger the number of hops, the larger the number of domains in sent for model inference, potentially, the larger the number of domains flagged as misinformation sources. 

For our experiments we restricted ourselves to our random forest models from section \ref{sec:model}. We decided to use the random forest technique since it performed the best across both versions of the model, binary and multi-class. For each experiment, we report the number of domains used as model input (All) and the number of domains identified as misinformation (or propaganda) by the models (Positive). We reported as a positive classification (i.e., the discovery of a misinformation domain) all domains that the model outputs a confidence score higher than 0.5 in all three months of data.

We present results for one-hop and two-hop egonoets in Table \ref{datavolume:binary}. We compared 1-hop and 2-hop egonet deployment and found that adding an egonet hop increases the number of domains sent for binary model classification from $\approx30$K to over 1.2 Million. The number of flagged domains increased by a factor of approximately two, from about 900 domains (one-hop) to about 2000 domains (two-hops).  In the multi-class analysis, the size of the set of domains selected for model input increases from $\approx4$K to over 1.2 Million without any increase in the number of findings.

Finally, we also compare the one-hop and two-hop deployment strategies with domains randomly sampled from traffic deployment. To estimate model performance in full traffic, we produced model predictions for 50k randomly selected domains. Afterward, we remove all domains with traffic lower than 3,000 visits a month from our initial 50k domains. Results for the sampled domains (after the removal of low traffic domains) are presented in Table \ref{datavolume:binary}. %describe the size of the set of domains in our experiments. For each month we count the number of domains used as input in each deployment strategy and the number of domains identified as misinformation (or propaganda) by the model. The number of domains used as input in the full traffic deployment varies because we exclude all domains with traffic lower than 3,000 visits a month.

\begin{table}
\centering
\begin{tabular}{l c c c c}

Month & Set & One-hop  & Two-hop & Sampled traffic   \\ \hline
   \multicolumn{5}{l}{Binary model: misinformation detection}\\ \hline
October  & All & 31,698 &  1,290,032 &  32,292 \\
 & Positive & 915 & 2,046 & 77\\
 
November & All & 31,130 & 1,277,717 & 32,251 \\
& Positive & 901 & 1,991 & 80 \\

   December & All & 31,779 & 1,191,569 & 31,779\\
 & Positive & 917 & 2,142 & 85 \\\hline
   \multicolumn{5}{l}{Multi-class model: propaganda detection}\\ \hline
  
October  & All & 4,494 &  1,289,231 &  32,292 \\
 & Positive & 57 & 57 & 2\\
 
November & All & 4,322 & 1,277,074 & 32,251 \\
& Positive & 44 & 44 & 2 \\

   December & All & 4,119 & 1,190.922 & 31,779\\
 & Positive & 47 & 45 & 2 \\ \hline

\end{tabular}
\caption{\label{datavolume:binary}Binary and multiclass model data volume for different deployment strategies. For each experiment, we report the number of domains used as model input (All) and the number of domains identified as misinformation (or propaganda) by the models (Positive).}\label{deploymentsize}
\end{table}

%For each domain in the deployment data set, we extract traffic features as described in section \ref{sec:feat}. We applied our random forest models from section \ref{sec:model} to the monthly datasets. For each experiment, we report the number of domains used as model input (All) and the number of domains identified as misinformation (or propaganda) by the models (Positive). We reported as a positive classification (i.e., the discovery of a misinformation domain) all domains that the model outputs a confidence score higher than 0.5 in all three months of data. Future efforts might loosen this restriction to better understand the overtime dynamics of misinformation sites or use the predictions from previous months as features.

%%% Revision 06/11

\subsection{Model Performance Assessment in Real Traffic Domains - The Binary Case}\label{dep:binary}

Given that our real traffic data set is unlabeled, we employ a strategy of manual verification for a random selection of domains to evaluate our model's precision and recall metrics. Our first deployment experiment measures the precision and recall of our binary model, described in section \ref{sec:model}. The chosen model for our deployment experiments was the random forest model -  the best-performing technique for the binary and multi-class cases. For our deployment experiments, we used the model trained on October data, and we computed features for the domains in the misinformation egonets present in October, November, and December.

\begin{table}
\centering
\begin{tabular}{l  c c }
\hline
   \multicolumn{3}{l}{Binary model: misinformation detection}\\ \hline
 Deployment  & Precision  & Recall  \\\hline
One-hop egonets & 0.78 & 0.81 \\
Two-hop egonets & 0.47 & 0.76 \\
Sampled traffic & 0.71 & 0.72 \\
\end{tabular}
\caption{\label{tab:deploydis} Binary model deployment results. We randomly sampled domains in all experiments to estimate precision and recall rates for each deployment strategy. Metrics in this table represent an average across all months. }
\end{table}

To estimate model precision, we randomly selected over 300 domains identified as misinformation by our model for each deployment strategy. Experts manually verified the domains, and results are presented as an average across three months. The process included 3 annotators. The annotator visits the page and selects 5 news articles to review from each section of the site. If at least three out of five news articles presented the following characteristics, the domain would be classified as misinformation: publish false content; engage in conspiracy theories; distort or misrepresent information.

Our model deployment estimated precision is 0.78 for the binary model when using the one-hop egonet deployment. The two-hop egonet deployment presented an estimated precision of 0.47. This means that in our one-hop egonet deployment, out of the 915 domains identified as misinformation (in October), about 714 domains are true positives. Analogously, for the two-hop deployment strategy out of the 2,046 domains identified as misinformation, about 961 domains are true positives. In the sampled traffic experiment, out of the 77 domains identified as misinformation, 54 domains were manually labeled as misinformation, resulting in a precision of 0.71. When comparing these three strategies, we see that the one-hop strategy has the highest signal-to-noise ratio, while providing a substantial number of findings, identifying over 700 domains (October).

We also selected 200 domains identified as non-misinformation by our model for each deployment strategy. We identified a very small number of domains that could potentially be misinformation. Scaling the data, we estimate that the recall for the one-hop egonet deployment strategy is around 0.8.

Compared to previous efforts \cite{hounsel2020identifying}, our framework produces outputs that can be easily consumed by content rating organizations, content moderators, and misinformation tracking efforts. In \cite{hounsel2020identifying} the authors report an estimated precision of 0.05. A likely factor that contributes to the low precision is the lack of a deployment strategy that can help address the inherent class imbalance between misinformation and non-misinformation sites. A similar work \cite{avast} reported the identification of 21 new misinformation domains when applied to real traffic. However, no precision analysis in deployment is presented in \cite{avast}. %Finally, \cite{chen2020proactive}

\subsection{The Multi-Class Case - Discovery of Previously Unknown Russian Propaganda Domains}\label{find:russian}

\begin{figure}
  \centering
  
 \includegraphics[width=5cm]{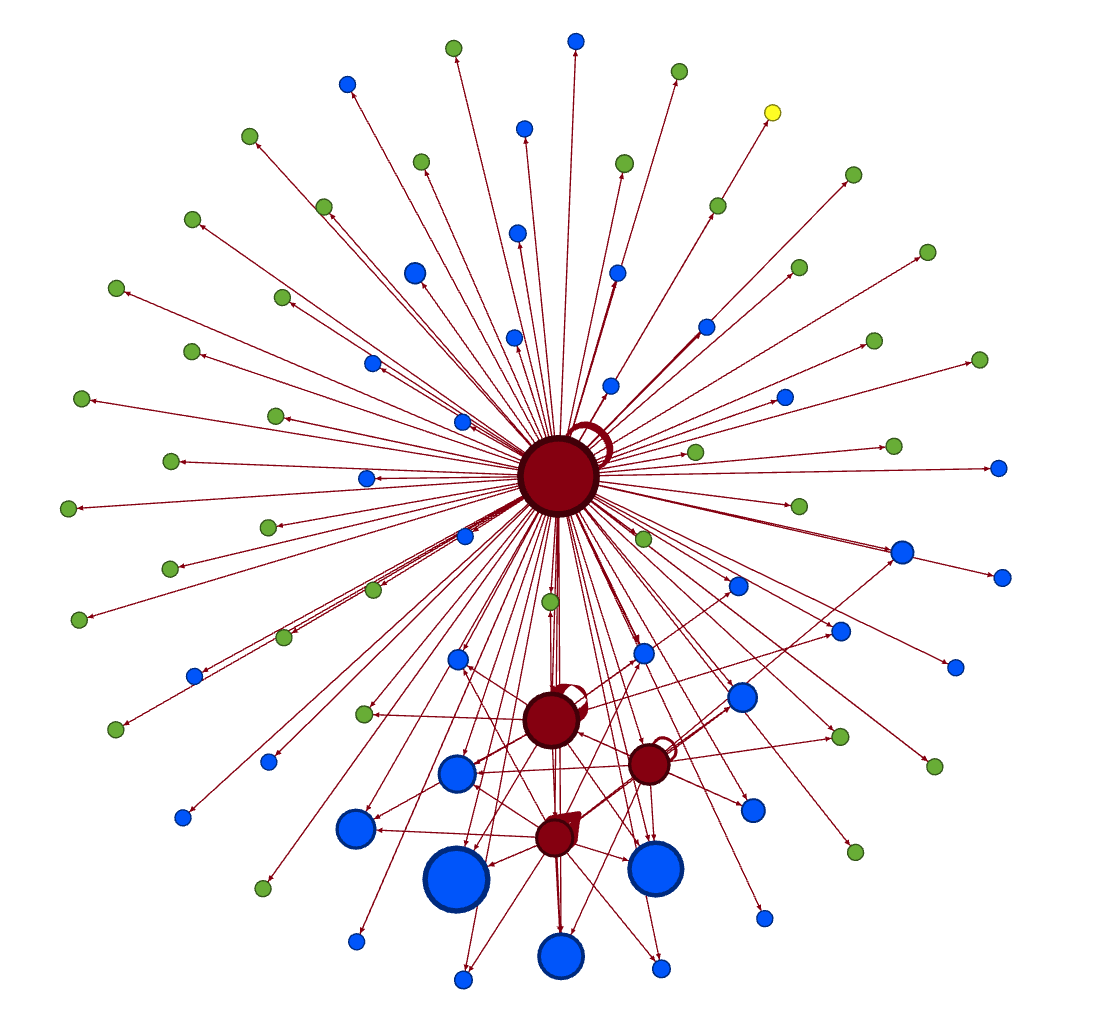}
  \caption{Outbound egonet of a Russian propaganda domain. Red nodes are Russian propaganda domains, yellow nodes are misinformation domains and blue nodes are domains labeled as non-misinformation. Green nodes are unlabeled domains.}\label{img:russianego}
\end{figure}

We deployed our multi-class model to identify Russian propaganda domains in three months of browser traffic data. The deployment process is similar to section \ref{dep:binary}. We started by selecting the unlabeled domains that are part of Russian propaganda domain's inbound and outbound egonets. Figure \ref{img:russianego} illustrates the outbound egonet of a Russian propaganda domain. In the graph, the red nodes are domains labeled as Russian propaganda, the blue nodes are non-misinformation domains, the yellow nodes are misinformation domains and the green nodes are unlabeled domains. We use our multi-class model to classify the green nodes, i.e. the unlabeled domains.

For three months of browser traffic, we analyzed the egonets of all Russian propaganda domains and extract the unlabeled domains with traffic greater than 3,000 visits a month.

In the one-hop deployment strategy, the model identified 57 domains as potential Russian propaganda spreaders in October, i.e. model confidence is greater than 0.5. All domains identified as propaganda by the model were reviewed by humans and 34 domains were confirmed as propaganda domains. The two-hop deployment strategy gave almost identical results. The full traffic experiments resulted in perfect precision, however, only two domains were identified by the model. We report in table \ref{tab:deployrp} averages across all months.

We also sampled domains that were identified as non-propaganda by the model to estimate recall. For each month, we reviewed 100 domains. Only one domain, in all three months, was identified as Russian propaganda.

\begin{table}
\centering
\begin{tabular}{l  c c }
\hline
   \multicolumn{3}{l}{Multi-class model: propaganda detection}\\ \hline
 Deployment  & Precision  & Recall  \\\hline
One-hop egonets & 0.56 & $\approx$0.99 \\
Two-hop egonets & 0.55 & $\approx$0.99 \\
Sampled traffic & 1.0 & $\approx$0.99 \\
\end{tabular}
\caption{\label{tab:deployrp} Multi-class model deployment results. Precision measures the ratio of true positives among the domains the model identified as propaganda spreaders.}
\end{table}

Compared to previous work, this is the first proposal of a multi-class model for identifying sub-categories of misinformation that has been effectively tested in a real deployment scenario. %This solution is particularly impactfull in scenarios where content rating agencies are interested in identifying domains spreading misinformation related to a specific theme, such as state-sponsored propaganda, and for-profit health scams, among others.

\section{Limitations and Ethical Considerations} \label{ethics}

\subsection{Traffic Data} Our model depends on browser traffic data to create traffic-based features, which may pose a limitation. Alternative data sources could include browser plugins or SEO tools \cite{panayotov2022greener}, which have been employed in similar contexts and could be adapted to our situation. Additionally, DNS traffic could be explored as another source of data.

\subsection{Bias} We understand the limitations of our labeled data, from the small size of the training data set, in which most non-misinformation websites are from the United States.  Additionally, we recognize the possibility of bias in our labels and judgments during the manual verification of results. To ensure a fair and unbiased process, we adhere to guidelines similar to those proposed by NewsGuard \cite{newsguard2021} for content verification. In the case of Russian propaganda verification, we specifically examined links, story duplication, and narrative similarities to known propaganda news portals, such as \textit{rt.com} and \textit{sputniknews.com}.

\subsection{Privacy} All data were collected with user consent. All unique identifiers (zip codes, rare domains) are removed. All PII is removed, and the authors do not have access to individual users’ data. Data is anonymized and aggregated. A similar approach to using browser data is found in \cite{greene2023rabbit}. Our data set only includes domains with more than 3,000 visits per month.

\subsection{Intended Use} This model is designed to help identify domains that may be spreading misinformation and propaganda. This model should be used in conjunction with human review, rather than as a standalone solution. When combined with human expertise, the model can greatly accelerate the process of detecting new misinformation domains.

\section{Conclusion}
% what we did 
% what we find 
% vague thing about the future

This study presents a framework for identifying misinformation domains using browser traffic features, drawing insights from both social and computer sciences. Our approach provides outputs that can be easily utilized by content rating organizations, content moderators, and misinformation tracking efforts.
%Our contributions are significant. 

Based on browser traffic patterns, we developed a method for accurately classifying domains as misinformation or non-misinformation. We also demonstrated that our approach is adaptable to multi-class scenarios, identifying subsets of misinformation domains, such as Russian state-sponsored propaganda. Finally, we proposed an efficient deployment framework to discover previously unknown misinformation domains. Our egonet-based strategy significantly increased the signal-to-noise ratio during deployment, achieving a precision over an order of magnitude greater than previous efforts \cite{hounsel2020identifying}.

Our work offers a practical solution for identifying misinformation domains with high precision rates, which can have significant implications for combating the spread of fake news and propaganda. As a  future research direction, we propose to explore the integration of page content into our predictive models.

{\footnotesize \bibliographystyle{acm}
\bibliography{sample}}

\end{document}